\documentclass[twocolumn,pra,showpacs,superscriptaddress]{revtex4}
\usepackage{graphicx}
\usepackage{amssymb,amsmath}
\usepackage{bm}
\usepackage{dcolumn}
\usepackage{subfigure}

\begin{document}

\title{Generating an effective magnetic lattice for cold atoms}

\author{Xinyu Luo}
\affiliation{State Key Laboratory of Low Dimensional Quantum Physics,
Department of Physics, Tsinghua University, Beijing 100084, China}
\affiliation{Institute of Physics, Chinese Academy of Sciences, Beijing 100080, China}
\affiliation{Collaborative Innovation Center of Quantum Matter, Beijing, China}

\author{Lingna Wu}
\author{Jiyao Chen}
\author{Rong Lu}
\affiliation{State Key Laboratory of Low Dimensional Quantum Physics,
Department of Physics, Tsinghua University, Beijing 100084, China}
\affiliation{Collaborative Innovation Center of Quantum Matter, Beijing, China}

\author{Ruquan Wang}
\affiliation{Institute of Physics, Chinese Academy of Sciences, Beijing 100080, China}
\affiliation{Collaborative Innovation Center of Quantum Matter, Beijing, China}
\author{L. You}
\affiliation{State Key Laboratory of Low Dimensional Quantum Physics,
Department of Physics, Tsinghua University, Beijing 100084, China}
\affiliation{Collaborative Innovation Center of Quantum Matter, Beijing, China}

\date{\today}

\begin{abstract}
We present a general scheme for synthesizing a spatially periodic
magnetic field, or a magnetic lattice (ML), for ultracold atoms using
pulsed gradient magnetic fields. Both the period and the depth of the artificial ML can be tuned,
immune to atomic spontaneous emission often encountered in optical lattices.
The effective Hamiltonian for our 2-dimensional ML has not been
discussed previously in condensed matter physics. Its band structures
show interesting features which can support topologically nontrivial phases.
The technical requirements for implementing our protocol
are readily available in today's cold atom experiments.
Realization of our proposal will significantly expand the repertoire
for quantum simulation with ultracold atoms.
\end{abstract}


\maketitle

Optical lattice (OL) is a highly controllable environment where many
body physics can be studied with ultracold atoms
\cite{Bloch-Review,hubbard toolbox}. Additionally, atoms in OL
promise exciting opportunities in quantum information science
\cite{1,2,3}. Many lattice geometries have been realized
experimentally, from 3-dimensional (3D)
cubic lattices \cite{bloch-2002} to honeycomb lattices
\cite{honeycomb} and kagome lattices \cite{kagome}. With
spin-dependent OLs, atomic internal degrees of freedom such as its
spin or pseudo-spin, are coupled to its spatial degrees of freedom.
This can give rise to interesting phenomena
\cite{vortex,Demarco,sengupta,bloch2,porto,sengstock2,Hof,demler,shankar,localization},
absent in spin-independent lattices. For example, attractive Fermi
gases in one-dimensional (1D) lattices support three-body bound
states with only two-body interactions when the tunneling rates are
spin-dependent \cite{luttinger-liquid}.
In 3-dimension, recent theoretical studies predict a exotic state with
the coexistence of superfluid and normal components \cite{vincent}.
For bosons, recent experiments reveal a new phase in spin-dependent OLs,
with one spin component
Mott-insulating and the other superfluid. The superfluid to Mott-insulating transition is modulated by their mutual interactions \cite{sengstock}.

A topical area of intense research interest in ultracold atoms
concerns synthetic gauge fields \cite{artificial gauge field}.
Several theoretical studies have proposed ideas to synthesize
artificial gauge potentials for atoms (with hyperfine spin $F$) in
OL systems \cite{zoller,dalibard,cooper1,cooper2,cooper3}, many
starting with simple forms $\propto F_z$ or spin-dependent lattices
and periodically driving the systems \cite{periodic1,periodic2,periodic3}.
Some of the ideas are realized in recent experiments
\cite{BlochPRL2013,KetterlePRL2013}. They emulate atomic
interactions with synthetic magnetic fields or spin-orbit coupling
(SOC). Additional interactions are therefore required to flip atomic
spins in order to broaden the scope of quantum simulations. More
general spin-dependent lattices and artificial gauge fields can
support exotic quantum
phases\cite{TRI,ZhangArxiv2012,TrivediPRL2012,GalitskiPRL2012,WuPRA2012}.
For example, W. Hofstetter {\it et al.} \cite{TRI} find fermionic
systems exhibit quantum phases such as topological and normal
insulator, metal, or semi-metal, all with two or more Dirac cones
even in the absence of atomic interactions when a staggered
potential is added to an artificial Rashba-type SOC. In the presence
of strong atomic interactions, semi-metal to antiferromagnetic
insulator transition can occur.

Spin-dependent OL can be readily generated by light shifts with
spin-dependent modulations \cite{GrynbergPR2001}, as in the familiar
lin-$\theta$-lin setup \cite{bloch2,PGC}. 1D effective Zeeman
lattice can be produced by combining a radio-frequency (RF) magnetic field with Raman laser fields \cite{GarciaPRL2012}.
However, due to the same scaling with laser intensity and detuning,
the ratio of spontaneous emission rate to the spin dependent lattice depth is determined by the ratio of atomic natural linewidth to its excite state fine structure
splitting. The resulting spin dependent lattice depth thus
 will be small in order to suppress spontaneous emission \cite{NovelLattice,KetterleArxiv2013,ZwierleinPRL2012}. Furthermore,
the ideas based on light-atom interaction are
limited by the laser wavelength, which make the spatial periods of the
resulting OLs difficult to tune. Larger spacing spin dependent lattice potentials can be realized
through microfabricated wires or permanent magnet arrays on an atom chip \cite{ZimmermannRMP2007,HannafordJPB2008,SpreeuwNJP2009}.

This work presents a different approach for synthesizing a spatially
periodic magnetic field or a magnetic lattice (ML) using
pulsed gradient magnetic fields. It can be understood in
terms of spatial dependent spin rotations, which couple atomic internal
degrees of freedom with its spatial/orbital degrees of freedom.
While sharing some features with the earlier mechanism
for SOC \cite{soc} and an analogous scheme using zero average modulated
gradient magnetic fields \cite{AndersonSOC},
the present idea opens the door towards a class of synthesized ML not previously explored.
It can be implemented by introducing a bias magnetic field to
the free evolution part of the SOC protocol \cite{soc}.
The ML lattice constant is tunable
and can overcome the laser wavelength limit.
Furthermore, the scheme we present can be generalized
in a straightforward manner to more than one spatial dimension.

\begin{figure}[tpb]
  \centering
  \subfigure{\includegraphics[width=\columnwidth]{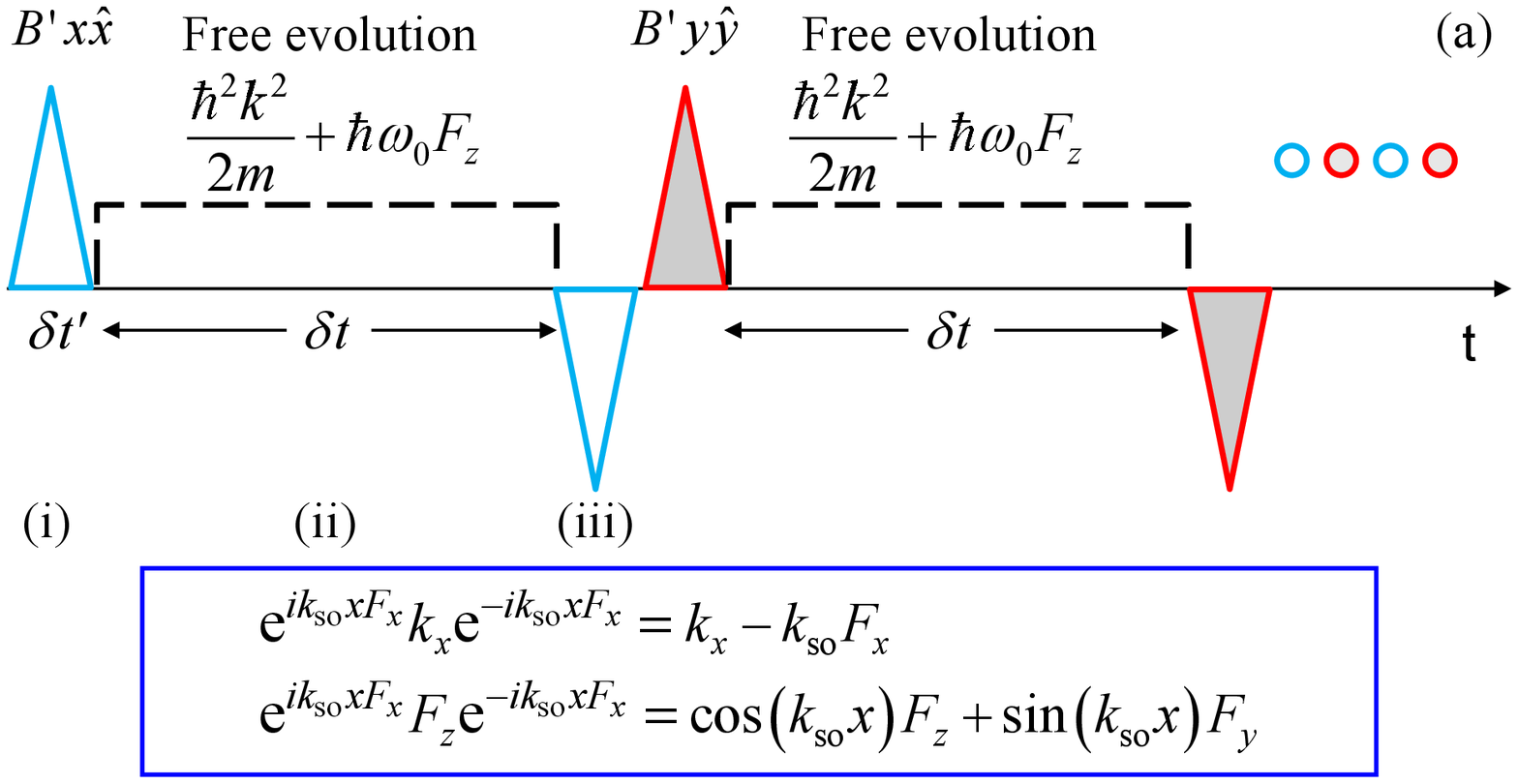}}
  \subfigure{\includegraphics[width=\columnwidth]{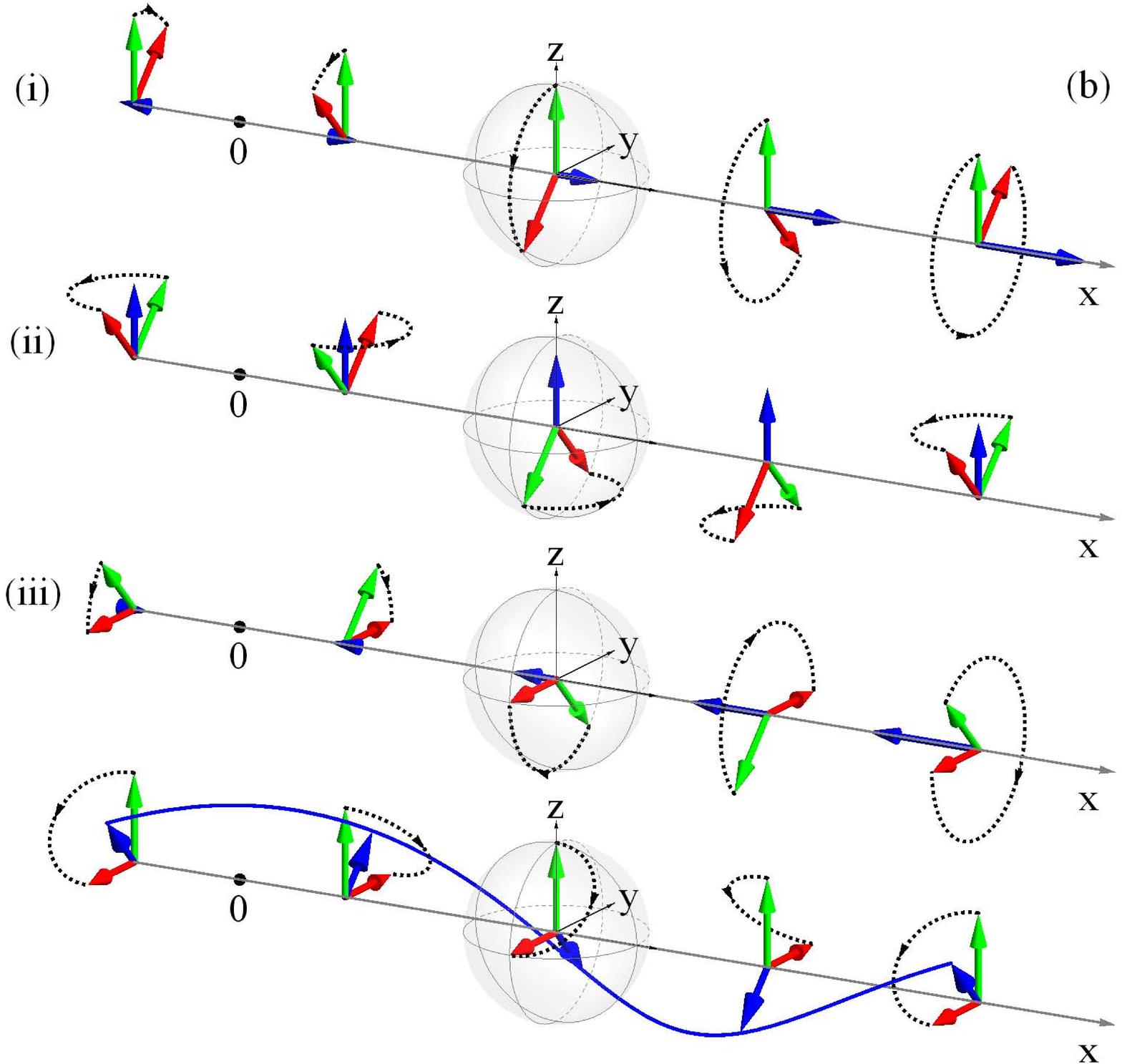}}
\caption{(a) A single period of our protocol starting with
free evolution in a bias field (in black
dashed line) sandwiched in between
a $x$-gradient pulse pair (in blue solid lines), followed by an
analogous $y$-gradient pulse pair (in red solid lines).
The opposite pulse pair affects unitary transformations as
displayed inside the rectangular box;
(b) Spatial dependent spin rotations are illustrated
for atoms located along the $x$-axis. Each row refers to
a different temporal instant marked in (a):
(i) [(iii)] at the first [second]
$x$-gradient pulse, affects clockwise [counter-clockwise]
spin rotations around the $x$-axis, while
free evolution in the $z$-bias field is shown in (ii).
In (i), (ii), and (iii),
green, red, and blue arrows denote respectively the initial,
the final spin directions, and the local magnetic fields.
The last row shows the synthesized ML (blue arrows), which is equivalent to
have spins rotated from the initial directions (green arrows)
to the final directions (red arrows).
The envelope for the ML is outlined in thick blue line.}
\label{fig1} 
\end{figure}

This Letter is organized as follows. First, our idea for
synthesizing a ML with a bias magnetic field
during the free evolution period of the SOC protocol \cite{soc}
is introduced. The dynamics governed by the synthesized Hamiltonian
are analyzed and numerically simulated, which support our claim that
the protocol we present is valid and effective.
The band structures for our ML are then computed.
We discuss conditions and signatures for experimentally
synthesizing and testing our ML.
Finally we summarize and discuss several
potential experimental challenges for implementing our idea.

Our idea for generating a 2-dimentional (2D) ML can be most easily
appreciated in comparison to the earlier SOC protocol using gradient
magnetic field pulses \cite{soc}. As is illustrated in Fig.
\ref{fig1}(a) for one period,
 the first (second) half is composed of free evolution in
a uniform magnetic field $B_0{\hat z}$, sandwiched in between two
short $x$- ($y$-) gradient magnetic field pulses $B'x{\hat x}$
($B'y{\hat y}$) with opposite amplitudes. $B'$ denotes the averaged
first order spatial derivative, or the spatial gradient, of the magnetic
field. $\delta t'$ denotes the duration of each pulse, which is
assumed small and the same for all pulses, while $\delta t$ is the duration
between the two pulses. $T/2=\delta t$ is half the period. Extending
the SOC protocol \cite{soc}, a nonzero bias magnetic field along
z-direction gives rise to the 2D ML.

\begin{figure}
\includegraphics[width=0.925\columnwidth]{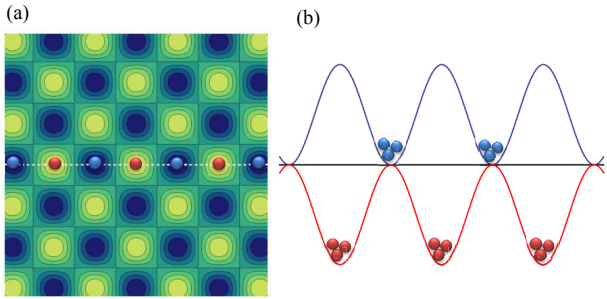}\\
\caption{(a) Internal state dependent lattice potentials for atoms in
the 2D ML;
(b) A 1D cut along the dashed line in (a). The blue and red lines
denote $M_F=1$ and $-1$ states, respectively. The magnetic field insensitive
$M_F=0$ state is denoted by the straight black line.
}
\label{fig2}
\end{figure}

\begin{figure}[tpb]
  \centering
  \subfigure{\includegraphics[width=0.49\columnwidth]{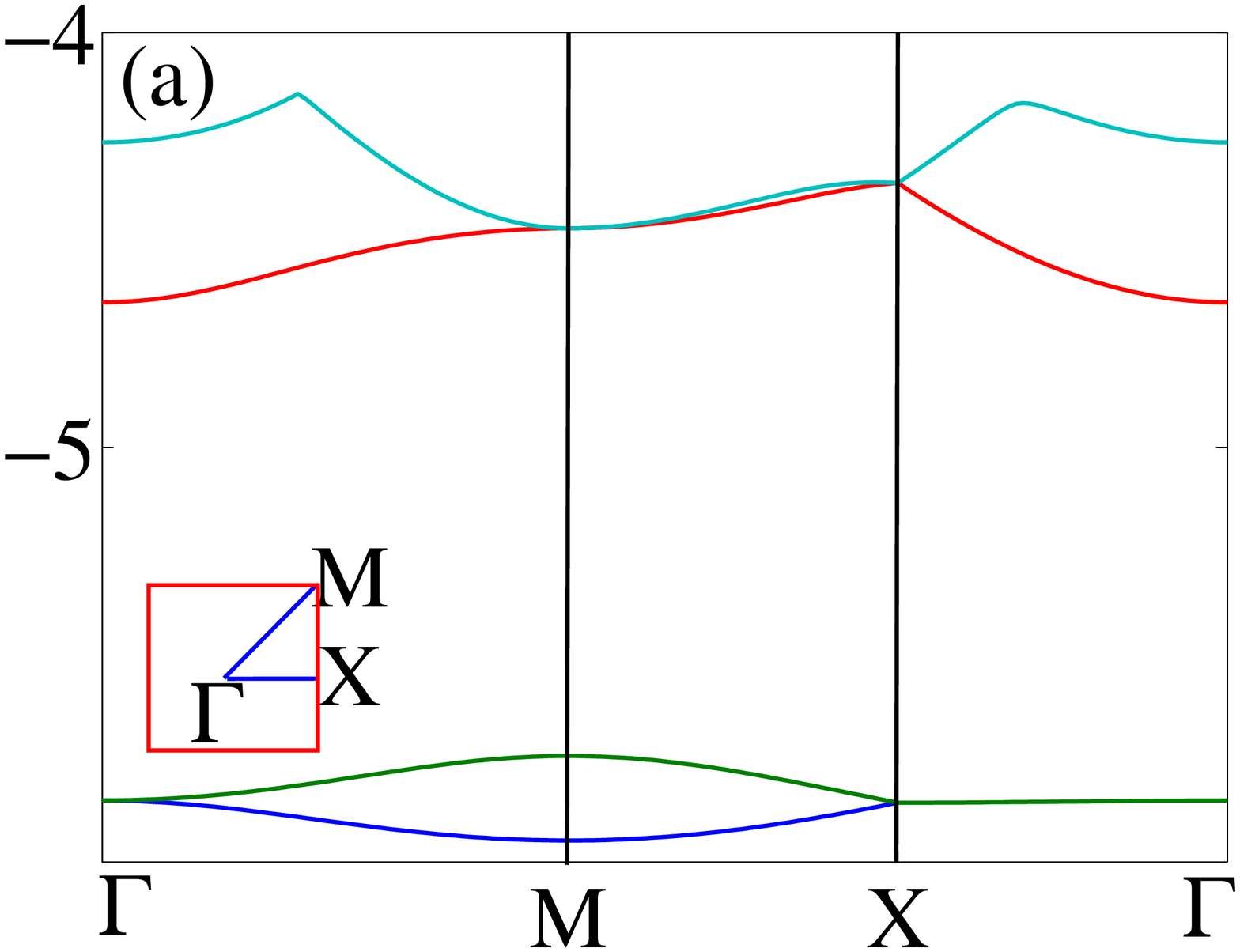}}
  \subfigure{\includegraphics[width=0.48\columnwidth]{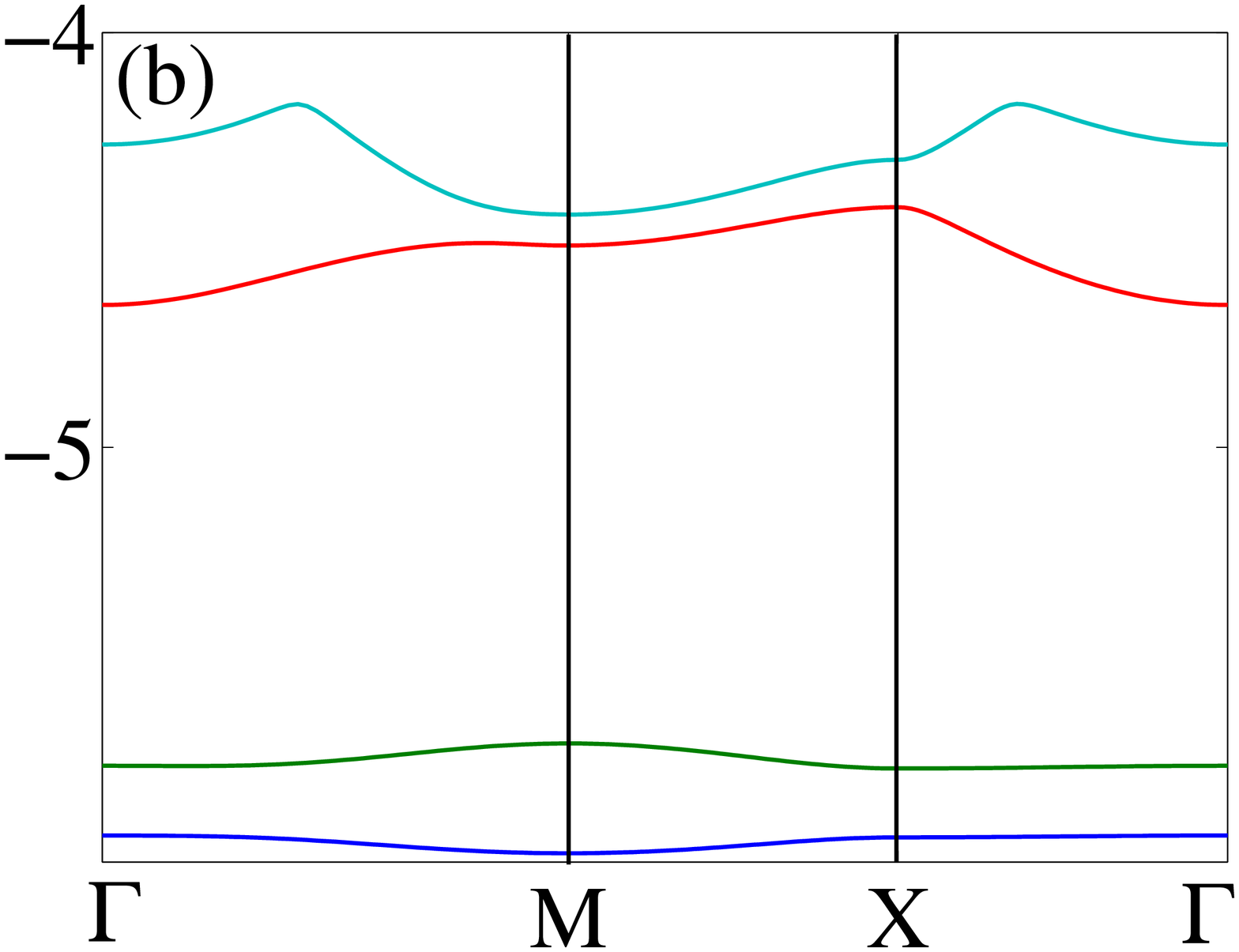}}
\caption{Single particle energy spectrum (in units of $\hbar\omega_R$)
for the lowest four bands, plotted for a contour in the
first Brillouin zone, as shown by the (red square) insert,
connecting reciprocal momentum points along  $\Gamma\to M \to X \to \Gamma$.
(a) At $\hbar\omega_0 =8\hbar\omega_R$, the band minima of the lowest
band is at $M$ point.
The band touching points are not accidental
as they exist for a wide range of lattice depth.
(b) The band degeneracy is broken in the presence of a
Zeeman term ($B = 0.1\hbar\omega_R$ used here).}
\label{fig3} 
\end{figure}

The dynamics from the first pulse are simple, with its evolution operator
given approximately by
${\cal U}_x(\delta t') =\exp(-ig_F{\mu}_BB^{'}xF_x{\delta}t^{'}/\hbar)$,
only due to the averaged Zeeman term $g_F{\mu}_BB^{'}xF_x$,
which is assumed to overwhelm all other interactions.
${\mu}_B$ is the Bohr magneton.
$g_F$ is the Lande $g$-factor for the $F=1$ ground state Zeeman
manifold considered.
$F_{x,y,z}$ denotes the $x$-, $y$-, and $z$-component of $\vec F$.
The pair of $x$-gradient pulses give
\begin{eqnarray}
 U_x(T/2,0) &=&{\cal U}_x(\delta t')
 \exp(-iH_0{\delta}t/\hbar)
{\cal U}_x^\dag(\delta t')  \nonumber\\
 &=& \exp(-iH_{\rm eff}^{(x)}\delta t/\hbar),
\label{e1}
\end{eqnarray}
which transforms the free evolution Hamiltonian
$H_0={\hbar^{2}k^{2}}/(2m)+\hbar \omega_0 F_z$
into a SOC plus a ML
\begin{eqnarray}
H_{\rm eff}^{(x)} &=& \frac{\hbar^{2}}{2m}\Big[(k_x-k_{\rm so}F_x)^{2}+k_y^2\Big]+\nonumber\\
&& \hbar\omega_0\Big[\cos(k_{\rm so}x)F_z+\sin(k_{\rm so}x)F_y\Big],
\label{e2}
\end{eqnarray}
where $\omega_0=g_F{\mu}_BB_0/\hbar$ is the Larmor
frequency at $B_0$.
The first line of Eq. (\ref{e2})
corresponds to SOC \cite{soc}, whose strength is given by the momentum impulse
${\hbar}k_{\rm so}=g_F{\mu}_BB^{'}{\delta}t^{'}$ from the gradient
pulse \cite{soc}. The second line corresponds to a ML
$\propto \sin(k_{\rm so}x){\hat y} + \cos(k_{\rm so}x){\hat z}$
with wave vector $k_{\rm so}$ as illustrated in Fig. \ref{fig1}(b).
Follow up with a pair of $y$-gradient pulses as shown in Fig. \ref{fig1}(a),
we end up with the two dimensional (2D) version
\begin{eqnarray}
U_{2D}(T,0)=U_{y}(T,{T/2})U_{x}({T/2},0)\approx \exp(-iH_{\rm eff}^{(2D)}T/\hbar), \nonumber
\end{eqnarray}
with
\begin{eqnarray}
H_{\rm eff}^{(2D)}&=&\frac{\hbar^{2}}{2m}(k_x-\frac{1}{2}k_{\rm so}F_x)^{2}
+\frac{\hbar^2k_{\rm so}^2}{8m} F_x^2 + \nonumber\\
&&\frac{\hbar^{2}}{2m}(k_y-\frac{1}{2}k_{\rm so}F_y)^{2}
+\frac{\hbar^2k_{\rm so}^2}{8m}F_y^2 + \nonumber\\
&&\frac{1}{2}\hbar\omega_0 [F_z \cos{(k_{\rm so}x)}+F_y \sin{(k_{\rm so}x)}]+\nonumber\\
&&\frac{1}{2}\hbar\omega_0 [F_z \cos{(k_{\rm so}y)}-F_x\sin{(k_{\rm
so}y)} ],
\label{heff2d}
\end{eqnarray}
provided the effective action from each cycle is small such that
we can use Trotter expansion to the first order and combine the
non-commuting $x$- and $y$-dependent terms into the same exponent.
${\hbar^2k_{\rm so}^2}(F_x^2+F_y^2)/{8m}$ acts like a
quadratic Zeeman shift. The leading order correction to the
time evolution operator for the 2D ML is $\delta U^{(1)}_{2D}(T,0)\approx\max(\omega^2_0,
\omega^2_R)\mathcal{O}(\delta t^2)$ assuming $k_x,k_y \lesssim
k_{\rm so}$, where $\omega_R={\hbar k^2_{\rm so}}/{2m}$. When the Trotter expansion fails, one can
simply reduce free evolution time and build up the action through
repeated pulse cycles \cite{soc}. More details can be found
in the supplemental material.
The synthesized $\vec B(x,y)$=$B_0[-\sin(k_{\rm so}y),
\sin(k_{\rm so}x),\cos(k_{\rm so}x)+\cos(k_{\rm so}y)]/2$ in our 2D ML
$H_{\rm eff}^{(2D)}$ gives the following eigenvalues $(|\vec B|,0,-|\vec B|) $,
as shown in Fig. \ref{fig2}(a).
Its typical band structure is shown in Fig. \ref{fig3}(a),
which is similar to lattice models employed for p-orbital physics \cite{xxx},
and is different from usual OLs. The lowest two bands touch at
$\Gamma$ $(K_x=0,K_y=0)$ and $X$ point $(k_{\rm so}/2,0)$ points.
The third and fourth bands touch at $M$ $(k_{\rm so}/2,k_{\rm so}/2)$ and $X$ points.
The band touching points are robust against tuning of lattice depth.
The degeneracy can be broken leading to a
gap if a Zeeman term $(\propto BF_z)$ is added
to Eq. (\ref{heff2d}) as shown in Fig. \ref{fig3}(b).

\begin{figure}
\centering
\includegraphics[width=\columnwidth]{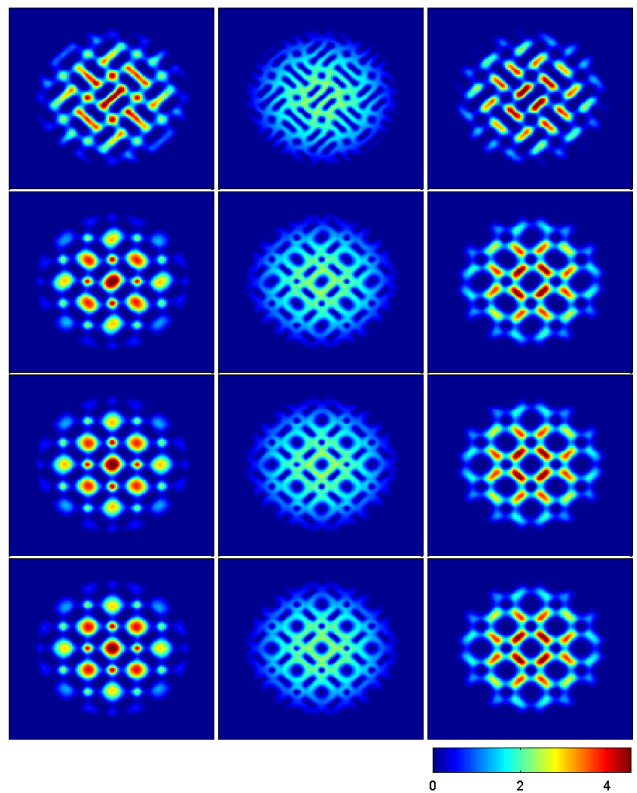}
\caption{The left, middle, and right columns respectively refer to
atomic internal states $|F=1,M_F\rangle$ for $^{87}$Rb with $M_F=1$, $0$, and $-1$.
Results from the actual dynamics are shown in the top 3 rows,
going down respectively for $\delta t=50$, $10$, and $2.5$ ${\mu}$s.
The last row comes from the effective dynamics.
$T=100$ $\mu$s, $B_0=10$ mGauss or $\omega_0=(2 \pi) 7.024$ kHz.
The gradient pulses are the same with $B'=1000$ Gauss/cm and
$\delta t'=1$ ${\mu}$s, which gives $k_{\rm so}=0.44$ ${\mu}$m$^{-1}$, and $\omega_R=(2\pi)11.23$ Hz.
The regions displayed correspond to $-15a_{h}<x,y< 15a_{h}$,
with ${a_{h}} = \sqrt {\hbar /m\omega }$ the length scale for an isotropic
harmonic oscillator $\omega_x=\omega_y=\omega=(2\pi)30$ Hz.}
\label{fig4}
\end{figure}

Figure \ref{fig4} compares a sampling of numerical simulations based on Gross-Pitaevskii equation.
For a fixed evolution time $T$,
actual dynamics are propagated using 1, 5, and 20 pulses
as shown in the 1st, 2nd, and 3rd row respectively.
With increasing numbers of pulse cycles,
atomic density distributions converge towards that
from the effective dynamics by $H_{\rm eff}^{(2D)}$,
which are shown in the last row.
We find that the required error bounds $\delta U^{(1)}_{2D}(T,0)\leq\max(\omega^2_0,
\omega^2_R)\delta t^2$ are always satisfied when $\delta t$ becomes sufficiently short.
Extensive simulations show that our idea is
effective and efficient. This conclusion is also supported
by the analytic derivation of the effective interaction above,
where all approximations used are reasonable under most circumstances.

The ML we synthesize can be straightforwardly detected
making use of lattice induced atomic diffractions as we illustrate
in detail in the supplemental material for the 1D case.
Both the SOC strength $k_{\rm so}$, or the reciprocal vector of the ML,
and the ML depth $\hbar\omega_0$ can be independently tuned in our protocol.
In the optical Raman scheme \cite{GarciaPRL2012,RamanSOC},
$k_{\rm so}$ is limited to two photon recoil momentum.
In our protocol, it is replaced by the momentum impulse from a
single gradient pulse, provided the kinetic energy term during the gradient pulse
can be neglected when $\omega_R
\delta t^{'}\ll1$, (see supplemental material). To achieve $k_{\rm so}=1\,(k_L)\sim 8$ ${\mu}$m$^{-1}$ (for a
typical laser wavelength $\lambda_L=804$ nm \cite{RamanSOC}), the
validity of our proposal requires $\delta t^{'}\ll 40$ $\mu$s and
$B^{'}\gg 427$ {Gauss/cm}. These conditions are already available in
atom chip experiments with pulses as short as 5 $\mu$s at a gradient
of 3.4 kGauss/cm \cite{FGBS}. One can even go beyond $k_{\rm so}\sim
k_L$. For $k_{\rm so}=4 k_L\sim 32$ $\mu$m$^{-1}$, $\delta
t^{'}\ll 2.5 $ ${\mu}$s is required if $B^{'}\gg 27.3$ kGauss/cm can
be generated. Such a short and high gradient pulse is of course an
experimental challenge. Nevertheless it is available in the state of
art NMR experiments (1.4 $\mu$s and 53.4 kGauss/cm)
\cite{shortpulse}, and is perhaps also realizable with improved atom
chips (0.5 $\mu$s and 170 kGauss/cm) \cite{FGBS} using thin wires
and tiny current loops. An unwelcomed drawback of using such high
gradient magnetic field is due to quadratic Zeeman shift, which
compromises energy level symmetry and reduces a spin $F$ atomic
system to a two-level or spin 1/2 system. It is interesting to point out
that our protocol for synthesizing 1D ML can be
alternatively viewed as a Ramsey like interferometer \cite{Ramsey}
with pulsed gradient magnetic fields acting as spatial dependent
oscillating fields. The population oscillations discussed in
the supplemental material are simply the interference signals.
In this sense, our protocol for 1D
ML also resembles the work of field gradient beam splitter (FGBS)
\cite{FGBS}.

The interaction terms $\propto F_x, F_y$ of our ML couple
spatial diffraction with atomic spin flips,
thus it can also be viewed as a particular type of SOC.
However, when reduced to 1D, the diffraction orders of our ML becomes finite,
constrained by the finite spin, rather than being infinite
as for the usual spin dependent lattices $\propto F_z$.
When spin flip pulses or additional free evolution periods are introduced
between successive 1D ML pulse pairs, the $F_y$ terms can be compensated for.
This will effectively reduce our ML in 1D to the usual $\propto F_z$
type lattice with broken continuous translation symmetry, allowing for spatial
diffraction to infinite orders.
The 2D ML synthesized with our protocol always breaks continuous
translation symmetry, thus is capable of infinite
order spatial diffraction by itself. It encompasses the combined effects of a usual spin dependent lattice and SOC.
Although the band structure we show earlier is topologically trivial,
several variants of Eq. (\ref{heff2d}),
generated by other pulse sequences, indeed display nontrivial topological
band structure, which will be further explored and published elsewhere.
To our knowledge, this type of 2D ML has not been
discussed in solid state materials.

Before conclusion, we discuss two experimental insights
regarding the feasibility of realizing our ML protocol
within the limits of current experimental capabilities. First,
Our protocol makes use of gradient magnetic fields such as $B'x{\hat x}$ to
enact spatial dependent spin rotations. Such magnetic fields,
however, cannot simply exist because the divergences
and curls of magnetic fields vanish in free space.
In actual experiments, the gradient field direction
is slaved to the direction of the large bias field $B_{x0}$,
as in selecting the $x$-gradient from the 3D quadruple field
$(B'x+B_{x0}){\hat x}+B'y{\hat y}-2B'z{\hat z}$ \cite{MagneticTrap}.
One can also use a gradient RF magnetic field with a
strong bias field to serve as gradient magnetic field \cite{AndersonSOC}.
Second, the $x$- ($y$-) direction gradient field pulses
must be turned on and off abruptly as otherwise atomic spins
will adiabatically follow the time dependent magnetic field.
Experimentally the temporal rate of changing magnetic field
is limited by the coil inductances and by eddy currents.
Small coils boosted by high voltage
power supplies can help produce rapid changing magnetic fields.
The eddy current effects can be measured and compensated for
by pre-emphasis current driving as widely employed
in pulsed field gradients NMR experiments \cite{eddycomp}.
Instead of flipping the direction of the gradient magnetic field,
we can alternatively flip the direction of atomic spin
through rf coupling \cite{spinecho} or motion insensitive Raman transitions.

In conclusion, we propose an idea for generating a ML
using pulsed gradient magnetic fields.
The spatial dependent spin rotations from the gradient fields
couple atomic internal states with its spatial motion,
effectively synthesizing a ML.
Both the lattice constant and its depth are tunable
experimentally and can overcome the laser wavelength limit
encountered in optical schemes.
By applying $x$- and $y$-gradient fields successively and
with sufficiently short free evolution times,
the 1D ML protocol discussed above can be extended to realize a 2D ML.
Atomic diffractions from the 1D ML give rise to
population oscillations among spin-momentum states,
which are easily observable and can falsify the synthesized ML.
Finally, we discuss experimental approaches for implementing our
protocols in today's cold atom experiments.
The protocol we propose can be applied to both bosonic and fermionic atoms.
The band structures of the synthetic ML display desirable features
of interest to topological quantum matter research.

This work is supported by MOST 2013CB922002 and 2013CB922004 of the National Key Basic Research Program of China,
and by NSFC (No.~91121005, No. ~11274195, No.~11004116, and No.~11374176).

\end{document}


\title{Supplementary material: generating an effective magnetic lattice for cold atoms}
\author{Xinyu Luo}
\affiliation{State Key Laboratory of Low Dimensional Quantum Physics,
Department of Physics, Tsinghua University, Beijing 100084, China}
\affiliation{Institute of Physics, Chinese Academy of Sciences, Beijing 100080, China}
\affiliation{Collaborative Innovation Center of Quantum Matter, Beijing, China}

\author{Lingna Wu}
\author{Jiyao Chen}
\author{Rong Lu}
\affiliation{State Key Laboratory of Low Dimensional Quantum Physics,
Department of Physics, Tsinghua University, Beijing 100084, China}
\affiliation{Collaborative Innovation Center of Quantum Matter, Beijing, China}

\author{Ruquan Wang}
\affiliation{Institute of Physics, Chinese Academy of Sciences, Beijing 100080, China}
\affiliation{Collaborative Innovation Center of Quantum Matter, Beijing, China}

\author{Li You}
\affiliation{State Key Laboratory of Low Dimensional Quantum Physics,
Department of Physics, Tsinghua University, Beijing 100084, China}
\affiliation{Collaborative Innovation Center of Quantum Matter, Beijing, China}

\date{\today}
\maketitle
\section{The validity conditions for the effective Hamiltonian}
In the main text, the effective two dimensional magnetic lattice Hamiltonian is derived by employing two
approximations. First, the Zeeman term $k_{\rm{so}}x F_x$ is assumed to overwhelm all other interactions
during the gradient magnetic field pulse, and second, higher order terms are neglected when
non-commuting exponents $A$ and $B$ are combined into the same exponent according to
${e^{A\delta t}}{e^{B\delta t}} \simeq {e^{(A + B)\delta t}}$. This section provides
the validity conditions for these two approximations by making use of the Trotter expansion
to the first order,
\begin{eqnarray}\label{s2}
{e^{A\delta t}}{e^{B\delta t}} \simeq e^{(A + B)\delta t}e^{\left[ {A,B} \right]\delta t^2/ 2}.
\end{eqnarray}

\subsection{The condition for neglecting atomic motion during the gradient pulse }
\begin{figure}
\includegraphics[width=0.7\columnwidth]{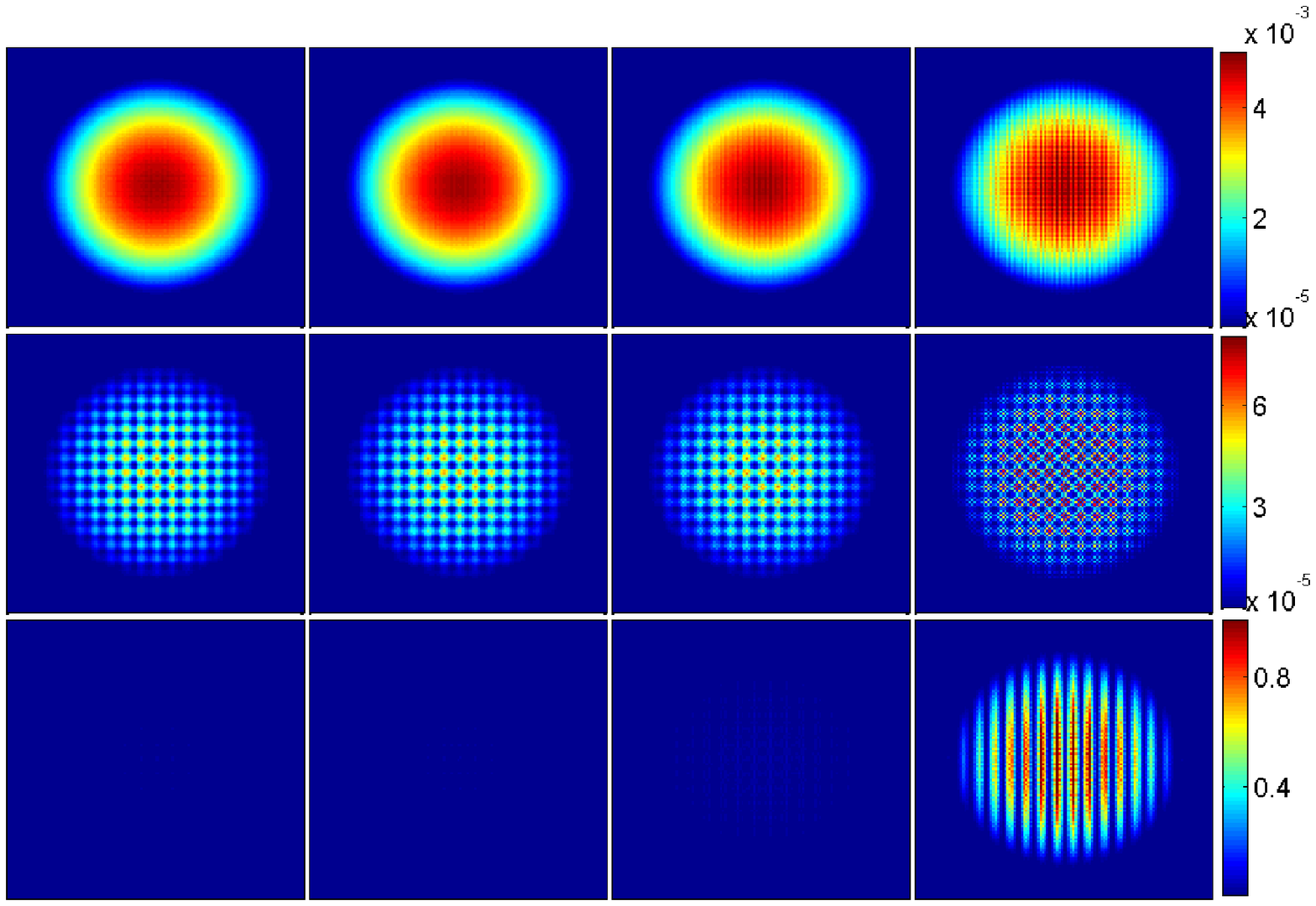}
\caption{Atomic density distributions from the effective Hamiltonian dynamics (in the first column)
are compared with results from the actual pulse sequence shown in the 2nd, 3rd, 4th columns,
respectively for $\omega_R \delta t^{'}=0.01$, $0.1$, and $1$. From top to down,
the three rows respectively show atomic densities in internal state
$|F=1,M_F\rangle$ for $^{87}$Rb with $M_F=1$, $0$, and $-1$.
The parameters used give the same momentum impulse $k_{\rm so}=1k_L \sim 7.8$ $\mu$m$^{-1}$,
(for a typical laser wavelength $\lambda_L = 804 n$m).
Other parameters are as discussed in the main text:
$\delta t=2.5$ $\mu$s, $B_0=10$ mGauss or $\omega_0=(2 \pi) 7.024$ kHz.
The regions displayed correspond to $-15a_{h}<x,y< 15a_{h}$,
with ${a_{h}} = \sqrt {\hbar /m\omega }$, the length scale for an isotropic
harmonic oscillator $\omega_x=\omega_y=\omega=(2\pi)30$ Hz.
}
\label{sm-fig1}
\end{figure}

Taking into account the atomic motion during the gradient magnetic pulse, the lowest order Trotter expansion for the single gradient
pulse evolution operator is found to be
\begin{eqnarray}\label{s6}
{\overline {\cal U} _x}(\delta t') &=& \exp \left\{ {i{k_{{\rm{so}}}}x{F_x} - i\frac{{\hbar {k_x}^2}}{{2m}}\delta t'} \right\} \notag \\
 &\approx& \exp \left\{ {i\frac{{\hbar {k_{{\rm{so}}}}}}{{2m}}{k_x}{F_x}\delta t'} \right\}\exp \left\{ { - i\frac{{\hbar {k_x}^2}}{{2m}}\delta t'} \right\}\exp \left\{ {i{k_{{\rm{so}}}}x{F_x}} \right\} \notag\\
 &=& \exp \left\{ {i{\omega _R}\delta t'{\left( {{{{k_x}} \mathord{\left/
 {\vphantom {{{k_x}} {{k_{{\rm{so}}}}}}} \right.
 \kern-\nulldelimiterspace} {{k_{{\rm{so}}}}}}} \right)}{F_x}} \right\}\exp \left\{ { - i{\omega _R}\delta t'{{\left( {{{{k_x}} \mathord{\left/
 {\vphantom {{{k_x}} {{k_{{\rm{so}}}}}}} \right.
 \kern-\nulldelimiterspace} {{k_{{\rm{so}}}}}}} \right)}^2}} \right\}\exp \left\{ {i{k_{{\rm{so}}}}x{F_x}} \right\}.
\end{eqnarray}
The last factor is the spatial dependent spin rotation discussed in the main text, while the deviations
of the first two multiplying from unity accounts for the error due to the neglect of atomic motion during
the gradient magnetic field pulse. The first factor denotes an `extra' spin rotation caused by atomic motion,
which can be safely neglected provided
\begin{eqnarray}\label{c2}
\omega_R
\delta t^{'}\ll1,
\end{eqnarray}
assuming $k_x,k_y \lesssim k_{\rm so}$. For ultracold atoms at tens or hundreds of nK
considered in this study, $k_x,k_y \lesssim k_{\rm so}$ is always satisfied
as it is simply equal to the statement of well below the recoil limit temperature,
if we take a more typical momentum impulse of ${k_{\rm so}}=1\left( {{k_L}} \right)$
The neglect of atomic motion during the gradient pulse can be
well satisfied as we illustrate
in Fig. \ref{sm-fig1}, which compares density distributions from the effective Hamiltonian
(in the first column) with that from the dynamics of the discussed pulse sequence
$\omega_R \delta t^{'}=0.01$, $0.1$, and $1$ (in the 2nd, 3rd, and 4th column, from left to right).
We see as $\omega_R \delta t^{'}$
becomes smaller, the approximation becomes better, and the
atomic density distributions uniformly converge towards that from the effective dynamics.

\subsection{The condition for achieving the 2D magnetic lattice}
We now proceed to investigate the condition for the second approximation.
According to our protocol developed in the main text, we set
\begin{eqnarray}\label{s3}
A &=&  \frac{\hbar }{{2m}}\left[ {{{\left( {{k_x} - {k_{\rm so}}{F_x}} \right)}^2} + {k_y}^2} \right] + {\omega _0}\left[ {{F_z}\cos \left( {{k_{\rm so}}x} \right) + {F_y}\sin \left( {{k_{\rm so}}x} \right)} \right] ,\nonumber\\
B &=&  \frac{\hbar }{{2m}}\left[ {{{\left( {{k_y} - {k_{\rm so}}{F_y}} \right)}^2} + {k_x}^2} \right] + {\omega _0}\left[ {{F_z}\cos \left( {{k_{\rm so}}y} \right) - {F_x}\sin \left( {{k_{\rm so}}y} \right)} \right],
\end{eqnarray}
the commutator between $A$ and $B$ is found to be
\begin{eqnarray}\label{s4}
\left[ {A,B} \right] &=& i \omega^2_R\left\{ {\left( {{k_x}/{k_{\rm so}} - {F_x}} \right),\left\{ {\left( {{k_y}/{k_{\rm so}} - {F_y}} \right),{F_z}} \right\}} \right\}+\nonumber\\
 &&i \omega _R\omega _0  \left[
\left\{ {\left( {{k_x}/{k_{\rm so}} - {F_x}} \right),{F_y}} \right\}\cos
\left( {ky} \right) + \left\{ {{k_y}/{k_{\rm so}},{F_z}\sin \left( {{k_{\rm so}}y}
\right){\rm{ + }}{F_x}\cos \left( {{k_{\rm so}}y} \right)} \right\}\right]+\nonumber\\
&&i\omega _R\omega _0 \left[ \left\{ {\left( {{k_y}/{k_{\rm so}} - {F_y}} \right),{F_x}} \right\}\cos \left( {kx} \right) - \left\{ {{k_x}/{k_{\rm so}},{F_z}\sin \left( {{k_{\rm so}}x} \right) - {F_y}\cos \left( {{k_{\rm so}}x} \right)} \right\}\right]+\nonumber\\
&&i\omega^2_0\left[ {{F_x}\sin \left( {{k_{\rm so}}x} \right)\cos \left( {{k_{\rm so}}y} \right) - {F_y}\cos \left( {{k_{\rm so}}x} \right)\sin \left( {{k_{\rm so}}y} \right) + {F_z}\sin \left( {{k_{\rm so}}x} \right)\sin \left( {{k_{\rm so}}y} \right)} \right],
\end{eqnarray}
where ${\omega _R} = \hbar{k_{\rm so}^2}/2m$, and $\left\{ {{F_i},{F_j}} \right\}= {F_i}{F_j}+{F_j}{F_i}$
denotes anticommutator. For spin-1/2 particles, $F_j=\sigma_j$,
which gives $\left\{ {{F_i},{F_j}} \right\} = 0$ if $i \ne j$. Equation (\ref{s4})
in the above reduces to
\begin{eqnarray}\label{s5}
[A,B] &=& i\omega^2_R\frac{{2{k_x}{k_y}}}{{{k_{\rm so}^2}}}{\sigma _z}+\nonumber\\
 &&i\frac{\omega _R\omega _0 }{{2k_{\rm so}}}\left[
\left\{ {{k_y},\cos \left( {k_{\rm so}x} \right) + \cos \left( {k_{\rm so}y} \right)} \right\}{\sigma _x}\right]+\nonumber\\
 &&i\frac{\omega _R\omega _0 }{{2k_{\rm so}}}\left[  \left\{ {{k_x},\cos \left( {k_{\rm so}x} \right) + \cos \left( {k_{\rm so}y} \right)} \right\}{\sigma _y} + \left( {\left\{ {{k_y},\sin \left( {k_{\rm so}y} \right)} \right\} - \left\{ {{k_x},\sin \left( {k_{\rm so}x} \right)} \right\}} \right){\sigma _z}
 \right]+\nonumber\\
 &&i\frac{\omega^2_0}{2}\left[ {{\sigma _x}\sin \left( {k_{\rm so}x} \right)\cos \left( {k_{\rm so}y} \right) - {\sigma _y}\cos \left( {k_{\rm so}x} \right)\sin \left( {k_{\rm so}y} \right) + {\sigma _z}\sin \left( {k_{\rm so}x} \right)\sin \left( {k_{\rm so}y} \right)} \right].
\end{eqnarray}
The leading order correction to the time evolution operator for the 2D ML
is therefore $\delta U_{2D}^{(1)}\left( {T,0} \right) \approx \max \left( {\omega _0^2,\omega _R^2} \right)O\left( {\delta {t^2}} \right)$. The condition for combining the otherwise non-commuting $x$- and $y$-dependent exponents
to forming the 2D ML in the same exponent is then simply given by
\begin{eqnarray}\label{c1}
\max \left( {\omega _0^2,\omega _R^2} \right)\delta {t^2} \ll 1,
\end{eqnarray}
which is supported by Fig. 4 in the main text.

\section{Detection of magnetic lattice}
A ML induces atomic diffraction, which in turn can be
used for its detection. We illustrate below for the 1D case.
In the parameter regime discussed here, $\omega_0\gg \omega_R$,
the Hamiltonian $H_{\rm eff}^{(x)}$ [Eq. (2) of the main text] then reduces to
\begin{eqnarray}
H_x=\hbar \omega_0 \left[F_z \cos({k_{\rm so} x})
+F_y \sin({k_{\rm so} x})\right],
\label{hx}
\end{eqnarray}
after neglecting the SOC term. It is invariant under continuous spatial translation
along $x$-direction. After a $\pi/2$ spin rotation
transformation about the $y$-axis, it becomes
\begin{eqnarray}
H_x^\prime &=& e^{ - i\frac{\pi }{2}F_y } H_x e^{i\frac{\pi }{2}F_y}  \nonumber\\
&=&  \frac{1}{2}\hbar \omega _0 \left( {F_ +e^{-ik_{\rm so} x}+F_-e^{ik_{\rm so} x} }\right),
\label{soc-1dp}
\end{eqnarray}
where $F_\pm =F_x\pm iF_y$ are the spin raising and lowering
operators. Thus the ML $H_x$ represents a
particular type of SOC, as an atom lowers (raises) its internal spin state,
it gains (or loses) momentum $\hbar k_{\rm so}$.
If the ML is suddenly turned on, atoms will start to
oscillate among different spin-momentum eigenstates $|M_F, k_x  =
nk_{\rm so}+k_x^{(0)} \rangle$ with frequency $\omega_0$,
similar to the Raman coupled SOC scheme. $k_x^{(0)}$ denotes the initial atomic kinetic
momentum. The spin-momentum population oscillation can be detected by time of fight Stern-Gerlach imaging,
which gives a clear signature of the synthesized 1D ML, as illustrated in Fig. {\ref{sm-fig2}.

\begin{figure}[!htb]
  \centering
  \subfigure{\includegraphics[width=0.247\columnwidth]{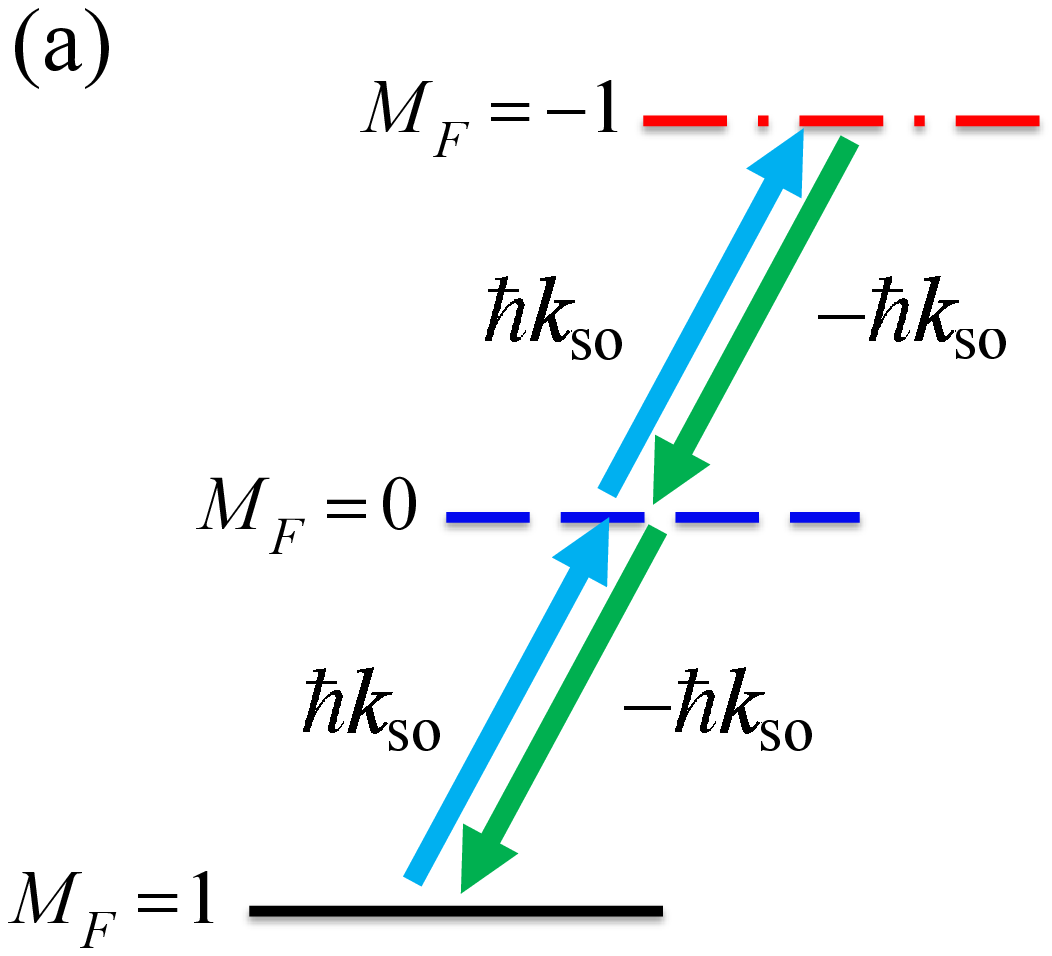}}
  \hspace{1.5cm}
  \subfigure{\includegraphics[width=0.39\columnwidth]{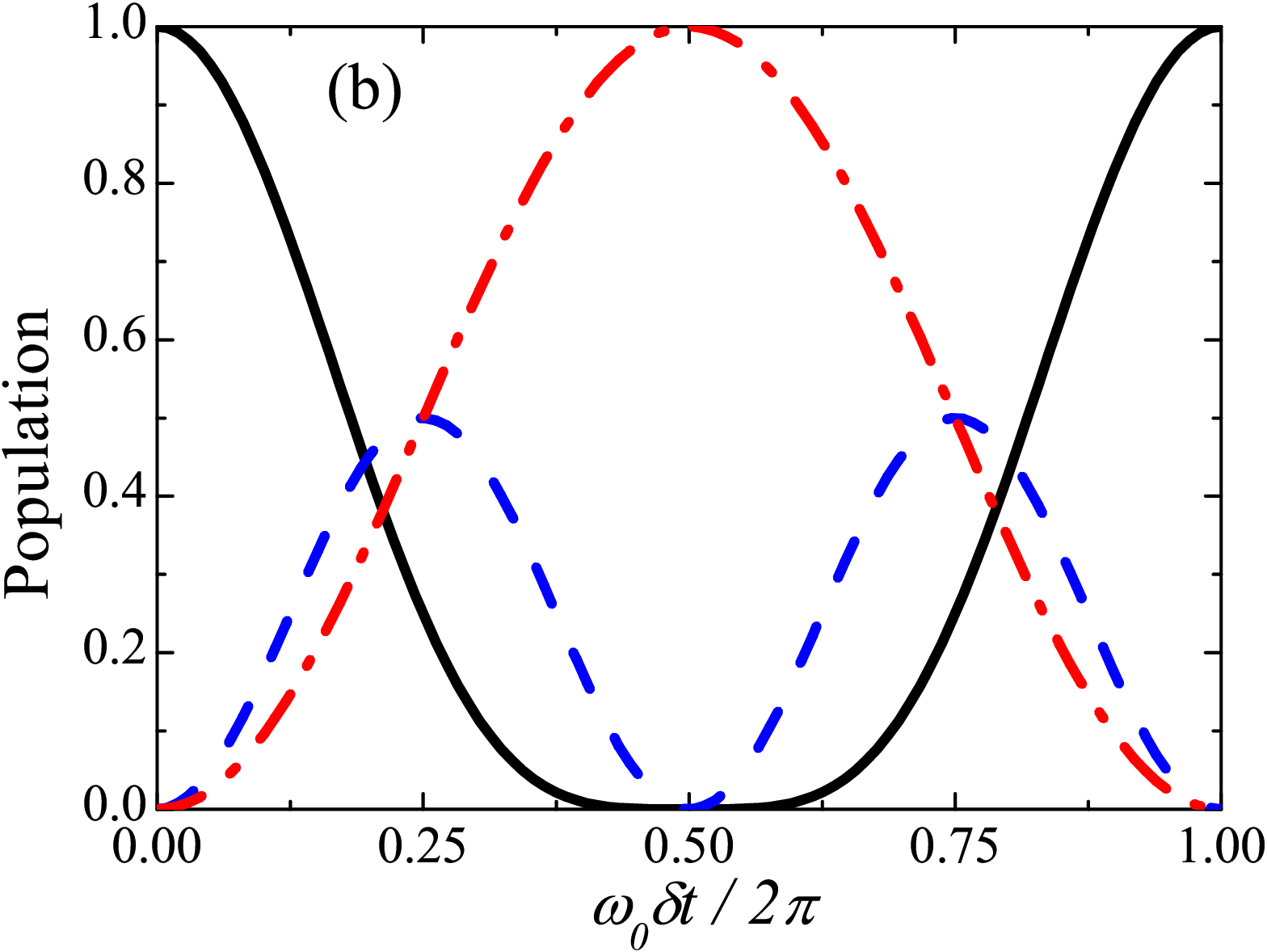}}
\caption{(a) The ML couples different spin-momentum states $|M_F, k_x  = nk_{\rm so} \rangle$.
Atoms gain (or lose) momentum $\hbar k_{\rm so}$ when their
internal spin states are lowered (raised).
(b) Oscillating populations.
The red dot-dashed, blue dashed, and black solid lines respectively
correspond to states $|M_F=-1,n=2\rangle$, $|M_F=0,n=1\rangle$, and $|M_F=1,n=0\rangle$.}
\label{sm-fig2} 
\end{figure}